\documentclass[preprint]{aastex}

\begin{document}

\title{Are the Kepler Near-Resonance Planet Pairs due to Tidal Dissipation?}

\author{Man Hoi Lee$^1$, D. Fabrycky$^{2,3,4}$, and D. N. C. Lin$^{3,5}$  
}
\affil{$^1$ Department of Earth Sciences and Department of Physics,
  The University of Hong Kong, Pokfulam Road, Hong Kong}
\affil{$^2$ Department of Astronomy and Astrophysics, University of
  Chicago, 5640 S.~Ellis Ave., Chicago, IL 60637, USA}
\affil{$^3$ UCO/Lick Observatory, University of California, Santa
  Cruz, CA 95064, USA}
\affil{$^4$ Hubble Fellow}
\affil{$^5$ Kavli Institute for Astronomy and Astrophysics and School
  of Physics, Peking University, China}

\begin{abstract}
The multiple-planet systems discovered by the Kepler mission show an
excess of planet pairs with period ratios just wide of exact
commensurability for first-order resonances like 2:1 and 3:2.
In principle, these planet pairs could have both resonance angles
associated with the resonance librating if the orbital eccentricities
are sufficiently small, because the width of first-order resonances
diverges in the limit of vanishingly small eccentricity.
We consider a widely-held scenario in which pairs of planets were
captured into first-order resonances by migration due to planet-disk
interactions, and subsequently became detached from the resonances,
due to tidal dissipation in the planets.
In the context of this scenario, we find a constraint on the ratio of
the planet's tidal dissipation function and Love number that implies
that some of the Kepler planets are likely solid.
However, tides are not strong enough to move many of the planet pairs
to the observed separations, suggesting that additional dissipative
processes are at play.
\end{abstract}

\section{INTRODUCTION}

The {\it Kepler} space telescope is designed to detect the periodic
transits of exoplanets in front of their host stars.
Based on the data obtained during the first 16 months of Kepler's
operation, more than 2000 planetary candidates have been identified,
analyzed, and published \citep{Batalha2013}.
Most of these planetary candidates have radii in the range 1--4 times
that of the Earth ($R_\oplus$) and orbital periods $P$ less than a few
months.
The rarity of planets with smaller radii and longer periods is due to
observational selection effects.
Although a large fraction of these candidates may indeed be planets,
some of them may be due to the blending of background eclipsing
binary stars with the light of foreground stars.  

\begin{figure}
\epsscale{0.8}
\plotone{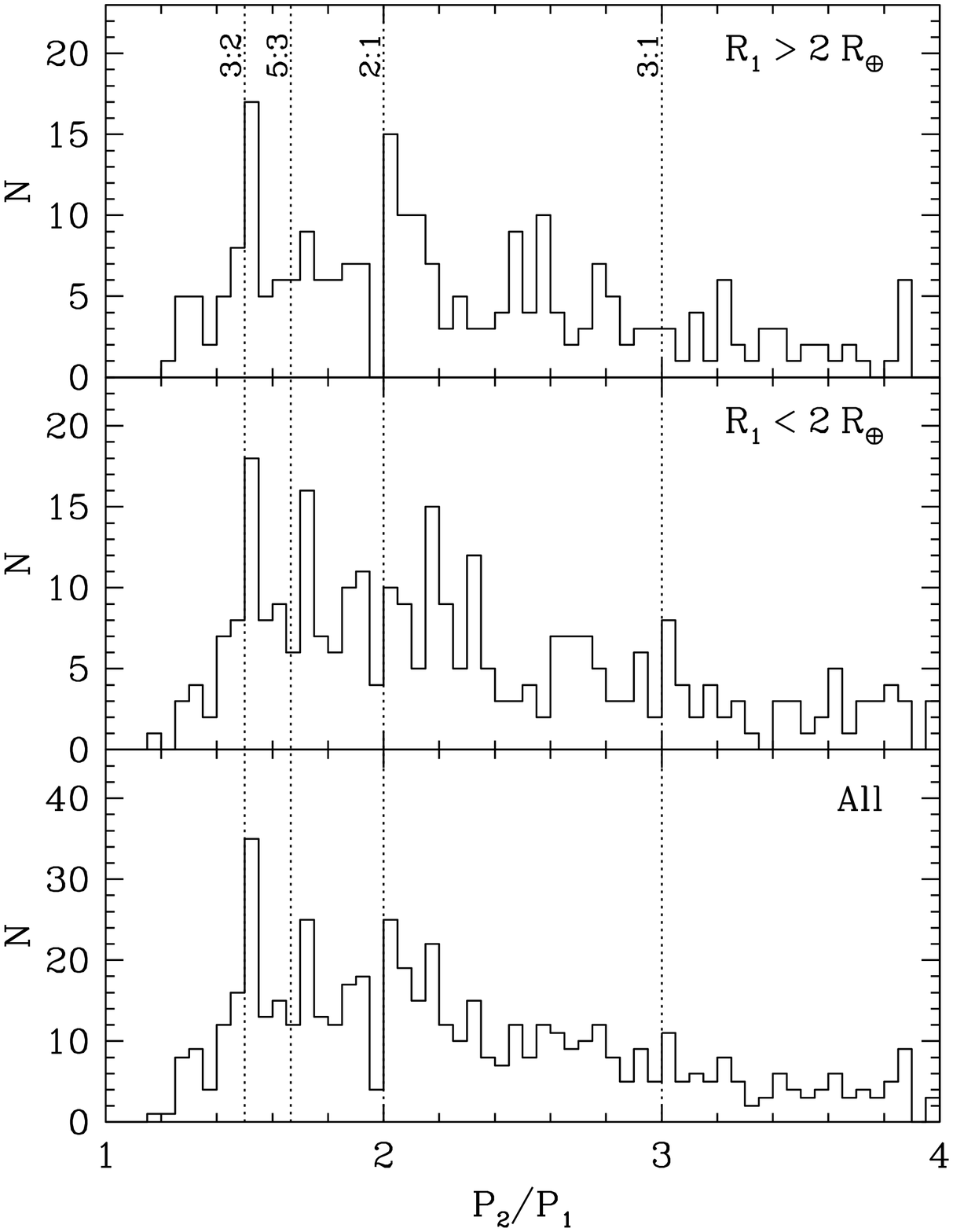}
\caption{
Histogram of period ratio $P_2/P_1$ for all Kepler candidate pairs
(bottom panel), pairs with radius of the inner planet $R_1 < 2
R_\oplus$ (middle panel), and pairs with $R_1 > 2 R_\oplus$ (upper
panel).
The dotted lines mark the exact period commensurabilities for the 3:1,
2:1, 5:3 and 3:2 resonances.
}
\label{fig:histogram}
\end{figure}

In the $16$-month Kepler catalog\footnote{
We did not use the 2-year catalog of \cite{Burke2013}, which is
available at http://exoplanetarchive.ipac.caltech.edu/, in this paper,
as the 2-year catalog was still changing, with evolving biases and
completeness, when we completed this work.
}, there are 361 host stars which bear
two or more transiting planetary candidates \citep{Fabrycky2012}, and
almost all of them are real multiple-planet systems
\citep{Lissauer2012}.
Their orbital configurations contain valuable information on the
history of their formation and dynamical evolution.
In particular, although a majority of the planet pairs are not in or
near mean-motion resonances, there is an excess of planet pairs with
outer-to-inner orbital period ratios, $P_2/P_1$, just wide of
first-order 2:1 and 3:2 resonances and a deficit of pairs with
$P_2/P_1$ just smaller than 2:1 and 3:2
\citep{Lissauer2011,Fabrycky2012}.
The excess and deficit occur within a few percent of exact
commensurabilities.
Figure \ref{fig:histogram} shows the histogram of period ratio for all
Kepler candidate pairs, pairs with radius of the inner planet $R_1 < 2
R_\oplus$ (Earths and super-Earths), and pairs with $R_1 > 2 R_\oplus$
(Neptunes and above).
Both pairs with $R_1 < 2 R_\oplus$, and pairs with $R_1 > 2 R_\oplus$
show an excess for $P_2/P_1$ just larger than 3:2 and a deficit just
smaller than 2:1, but interestingly, there is not an obvious excess
just larger than 2:1 for pairs with $R_1 < 2 R_\oplus$ (although the
statistics is noisy due to small numbers).
Also, the lack of pairs with $P_2/P_1$ just smaller than 3:2 reported
by \cite{Fabrycky2012} is not noticeable in Figure
\ref{fig:histogram}, because the bin size ($0.05$) is large compared
to the width of the gap ($0.01$--$0.02$).\footnote{
\cite{Fabrycky2012} have found that the distribution of period ratios
around second-order resonances is consistent with a random
distribution.
The excess of pairs with $R_1 < 2 R_\oplus$ (as well as all pairs) for
the bin in Figure \ref{fig:histogram} centered at $P_2/P_1 = 1.725$,
which is just wide of the second-order 5:3 resonance, is likely an
artifact, as there are no longer obvious features around 5:3 when we
shift the bins by, e.g., $0.02$ in $P_2/P_1$.
}

The Kepler candidates should be contrasted with the radial velocity
sample, which also shows an excess of planet pairs near the 2:1
resonance.
The radial-velocity planets are mostly Jupiter-mass planets, and the
excess near the 2:1 resonance consists of confirmed (or likely)
resonant pairs such as GJ~876
\citep{Laughlin2001,Rivera2001,Laughlin2005} and HD~82943
\citep{Lee2006,Tan2013}.
A widely accepted hypothesis for the origin of such resonances is
resonance capture through convergent migration of the planets
\citep{Bryden2000,Kley2000,Lee2002}.
Tidal interaction between an embedded planet and its natal
protoplanetary disk generally leads to a torque imbalance
\citep{Goldreich1980}.
Jupiter-mass planets are able to open gaps in the disks, and they
generally evolve with the viscous diffusion of their natal disks and
undergo inward type II migration (except in the outermost regions of
the disk where the disk spreads viscously outward and the migration is
outward) \citep{Lin1986}.
Convergent migration occurs if the inward migration of the outer
planet proceeds faster than that of the inner planet and the
outer-to-inner period ratio, $P_2/P_1$, decreases.

Lower-mass planets, like most of the Kepler candidates, do not
significantly perturb the disk surface density distribution, and thus
they undergo type I migration.
The direction of type I migration is expected to be inward in the
classic theory of, e.g., \cite{Ward1997}, but more recent analysis
have shown that type I migration could be outward in the viscously
heated regions of some disk models
\citep{Paardekooper2011,Kretke2012}.

Tidal interactions also occur between planets and their host stars,
which generally lead toward a state of spin synchronization and
orbital circularization (see, e.g., \citealt{Hut1981,Peale1999}).
If the host spins with a frequency faster (slower) than its
companion's orbital mean motion, the dissipation of the tides raised
by the companion in the host would reduce (increase) the host's spin
frequency and increase (decrease) the companion's orbital semimajor
axis on a timescale which is determined by the host's $Q/k_2$ value,
where $Q$ is the tidal dissipation function and $k_2$ is the Love number.
The companion's eccentricity is generally damped by the dissipation of
the tides raised by the host in the companion on a timescale which is
determined by the companion's $Q/k_2$ value.

In this paper, we consider in detail a scenario which may account for
the Kepler near-resonance planet pairs.
Pairs of planets were captured into first-order resonances by
convergent migration due to planet-disk interactions, and
subsequently became detached from the resonances due to tidal
dissipation in the planets.
The latter process was proposed by \cite{Novak2003} and more recently
by \cite{Papaloizou2010}, \cite{Lithwick2012a}, and \cite{Batygin2013}.
In Section 2, we give a simple explanation why the period ratio
$P_2/P_1$ should in fact be slightly larger than the exact period
ratio for a first-order resonance, if both resonance angles are
librating.
We then show that classic type I migration should result in resonant
pairs with $P_2/P_1$ much closer to exact commensurability than the
few percent departures observed for the Kepler near-resonance pairs.
Subsequent tidal damping of eccentricity and evolution away from exact
commensurability are discussed in Section 3, along with known
constraints on $Q/k_2$ for rocky and giant planets.
Comparisons with the observed Kepler near-resonance pairs are used to
derive constraints on $Q/k_2$ and the rocky/giant nature of the Kepler
planets in Section 4.
The results are summarized and discussed in Section 5.

\section{RESONANT PLANET PAIRS}

For first-order, $j$:$(j-1)$, mean-motion resonances like 2:1 and 3:2,
there are two eccentricity-type resonance angles,
\begin{eqnarray}
\theta_1 &=& (j-1) \lambda_1 - j \lambda_2 + \varpi_1 , \\
\theta_2 &=& (j-1) \lambda_1 - j \lambda_2 + \varpi_2 ,
\end{eqnarray}
where $\lambda_i$ is the mean longitude of planet $i$ and $\varpi_i$
is the longitude of periapse ($i = 1$ and $2$ for the inner and outer
planets, respectively).
At least one of these angles must librate about a fixed value for the
pair to be in resonance, if we ignore inclination resonances.
The resonance induced periapse precession is usually {\it retrograde},
i.e., $\dot\varpi_i < 0$ (see below).
In the best example of a mean-motion resonance in extrasolar planetary
systems, the 2:1 resonance in GJ 876, both angles librate about
$0^\circ$, and the periapses are observed to precess at an average
rate of $\dot\varpi_i = -41^\circ\,$yr$^{-1}$
\citep{Laughlin2001,Rivera2001,Lee2002,Laughlin2005}.
If the resonance angle $\theta_i$ is librating,
\begin{equation}
{\dot\theta_i} = (j-1) n_1 - j n_2 + {\dot\varpi_i} = 0
\end{equation}
on average, or
\begin{equation}
{P_2 \over P_1} - {j \over (j-1)} = - {{\dot\varpi_i} \over (j-1) n_2} > 0 ,
\label{eq:pratio}
\end{equation}
where $n_i$ is the mean motion, $P_i = 2\pi/n_i$ is the orbital
period, and $\dot\varpi_i < 0$ due to the resonance.
So $P_2/P_1$ should in fact be slightly larger than the exact period
ratio for a resonant pair.

In the vicinity of a $j$:$(j-1)$ resonance, the Hamiltonian to the
lowest order in the orbital eccentricities $e_i$ is
\begin{equation}
H = - {G M_\ast M_1 \over 2 a_1} - {G M_\ast M_2 \over 2 a_2}
    - {G M_1 M_2 \over a_2} (C_1 e_1 \cos\theta_1 + C_2 e_2 \cos\theta_2) ,
\end{equation}
where $M_\ast$ is the stellar mass, $M_i$ is the planetary mass, and
$a_i$ is the orbital semimajor axis (e.g.,
\citealt{Peale1986,Murray1999}).
The coefficients $C_1 = (1/2)(-2j - \alpha D) b^{(j)}_{1/2}$ and $C_2
= (1/2)(-1 +2j + \alpha D) b^{(j-1)}_{1/2} - \delta_{j2}/(2
\alpha^2)$, where $\alpha = a_1/a_2$, $b^{(j)}_{1/2}(\alpha)$ is the
Laplace coefficient, $D = d/d\alpha$, and $\delta_{j2}$ is the
Kronecker delta.
For 2:1, $C_1 = -1.190$ and $C_2 = 0.428$.
For 3:2, $C_1 = -2.025$ and $C_2 = 2.444$.
The first two terms of the Hamiltonian are the unperturbed Keplerian
Hamiltonian and the remaining terms are the resonant interactions.
The equation for the variation of the periapse longitude is
\begin{equation}
{d\varpi_i \over dt} = -{\sqrt{1 - e_i^2} \over M_i e_i \sqrt{G M_\ast a_i}}\,
                        {\partial H \over \partial e_j} ,
\label{eq:dwdt}
\end{equation}
if we assume coplanar orbits.
Since libration of $\theta_1$ about $0^\circ$ and $\theta_2$ about
$180^\circ$ is the only stable resonance configuration for small
eccentricities (e.g., \citealt{Peale1986,Murray1999,Lee2004}),
$\dot\varpi_1 = \alpha n_1 (M_2/M_\ast) C_1/e_1$ and $\dot\varpi_2 =
-n_2 (M_1/M_\ast) C_2/e_2$ to the lowest order in eccentricities.
Thus $\dot\varpi_i \propto -1/e_i$ and $P_2/P_1 - j/(j-1)$ could be
large and positive, if the eccentricities are small.
This is different from higher-order resonances.
For example, for a second-order resonance, the resonant terms in the
Hamiltonian to the lowest order in the eccentricities are proportional
to $e_1^2$, $e_1 e_2$, and $e_2^2$, and $\dot\varpi_1$ involves terms
that are either independent of the eccentricities or proportional to
$e_2/e_1$ (similarly, $e_1/e_2 $ for $\dot\varpi_2$).

We can derive two simple relationships between the eccentricities and
the period ratio from the above expressions for $\dot\varpi_1$ and
$\dot\varpi_2$.
From the requirement that both orbits precess at the same rate on
average, i.e., $\dot\varpi_1 = \dot\varpi_2$,
\begin{equation}
{e_2 \over e_1} = -\alpha^{1/2} {C_2 \over C_1} {M_1 \over M_2}
\label{eq:e2e1}
\end{equation}
\citep{Lee2004}.
Substituting $\dot\varpi_1$ into equation (\ref{eq:pratio}),
\begin{equation}
{P_2 \over P_1} - {j \over j-1} = {-1 \over (j-1)} \alpha^{-1/2}
{M_2 \over M_\ast} {C_1 \over e_1} .
\label{eq:pratioe1}
\end{equation}
In the above equations, $\alpha = a_1/a_2 \approx [(j-1)/j]^{2/3}$.

Convergent migration of planets due to interactions with the
protoplanetary disk can result in capture into mean-motion resonances.
This is the most likely scenario for the origin of the 2:1 resonance
in the GJ 876 system \citep{Lee2002}.
If the growth of eccentricity due to continued migration within the
resonance is balanced by the damping of eccentricity by planet-disk
interactions, the eccentricities would reach equilibrium values
determined by the ratio of the rates of eccentricity damping and
migration.
A natural question arises as to whether the Kepler near-resonance
pairs are simply resonance pairs with very small eccentricities (and
hence large positive departure of $P_2/P_1$ from exact
commensurability) due to large eccentricity damping during
disk-induced migration.\footnote{
At sufficiently small eccentricities, the regions of libration and
circulation of the resonant angles are not separated by separatrices,
and it is often said that the pair is no longer in resonance (e.g.,
\citealt{Delisle2012}).
However, our analysis applies as long as the interactions between the
planets are dominated by the libration of the resonant angles.
}
Most of the Kepler candidate planets are sufficiently small that they
are unable to open gaps in the protoplanetary disks and should undergo
type I migration.
For {\it classic} type I migration, the migration rate is
\citep{Ward1997,Tanaka2002}
\begin{equation}
{{\dot a} \over a} = - C_a {M_p \over M_\ast} {\Sigma a^2 \over
  M_\ast} \left(H \over a\right)^{-2} {2\pi \over P} ,
\label{eq:dadt}
\end{equation}
and the eccentricity damping rate is \citep{Artymowicz1993}
\begin{equation}
{{\dot e} \over e} = -9 C_e {M_p \over M_\ast} {\Sigma a^2 \over
  M_\ast} \left(H \over a\right)^{-4} {2\pi \over P} ,
\end{equation}
where $C_a \approx 3$, $C_e \approx 0.1$, $M_p$ is the planetary mass,
$\Sigma$ is the surface mass density of the disk, and $H/a$ is the
dimensionless scale height of the disk.
The ratio
\begin{equation}
K_e = \left|{{\dot e}/e} \over {\dot a}/a\right| = {9 C_e \over C_a}
\left(H \over a\right)^{-2} .
\end{equation}
For $H/a = 0.05$ and $0.1$, $K_e = 120$ and $30$, respectively.

\begin{figure}
\epsscale{1.1}
\plottwo{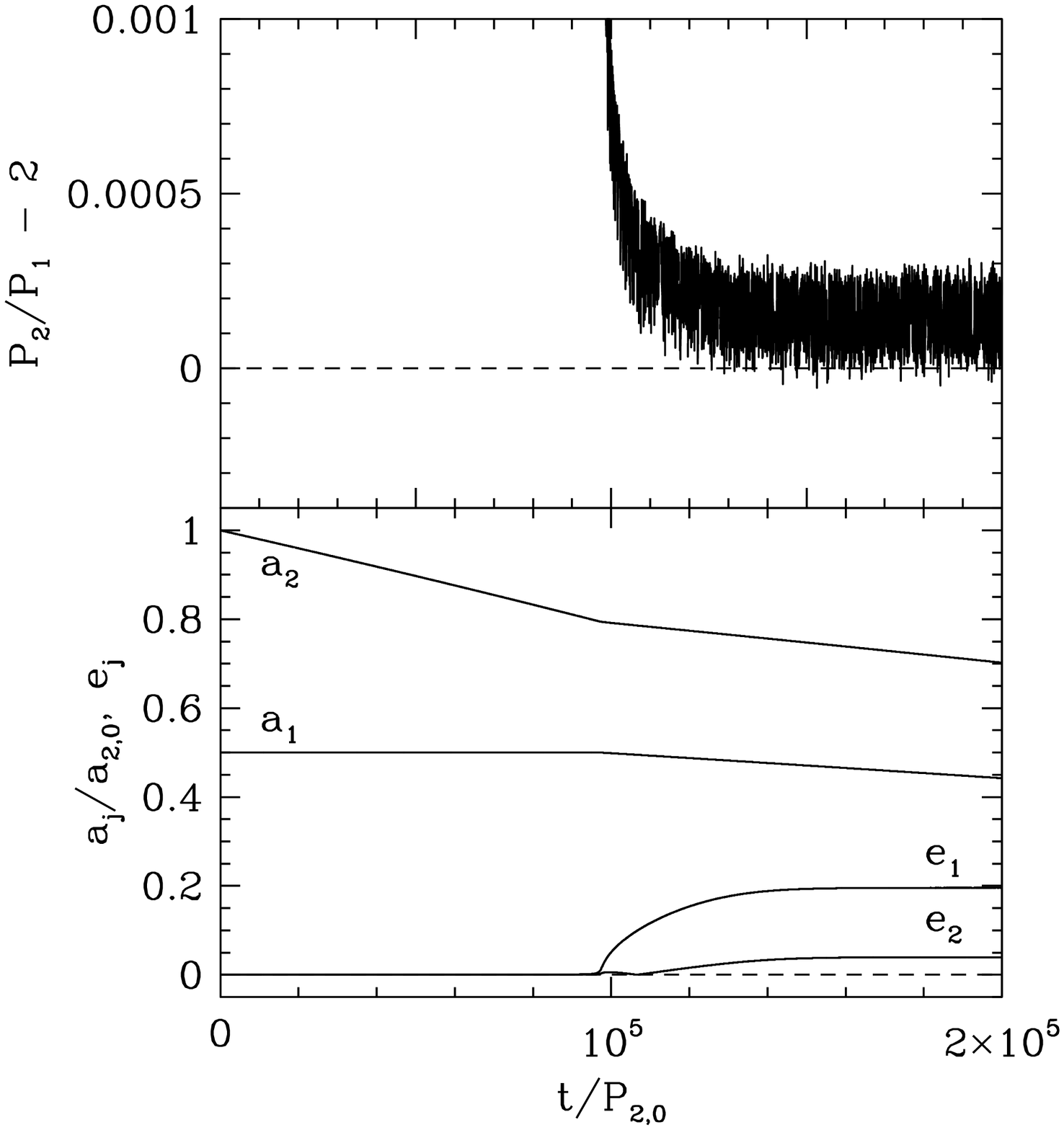}{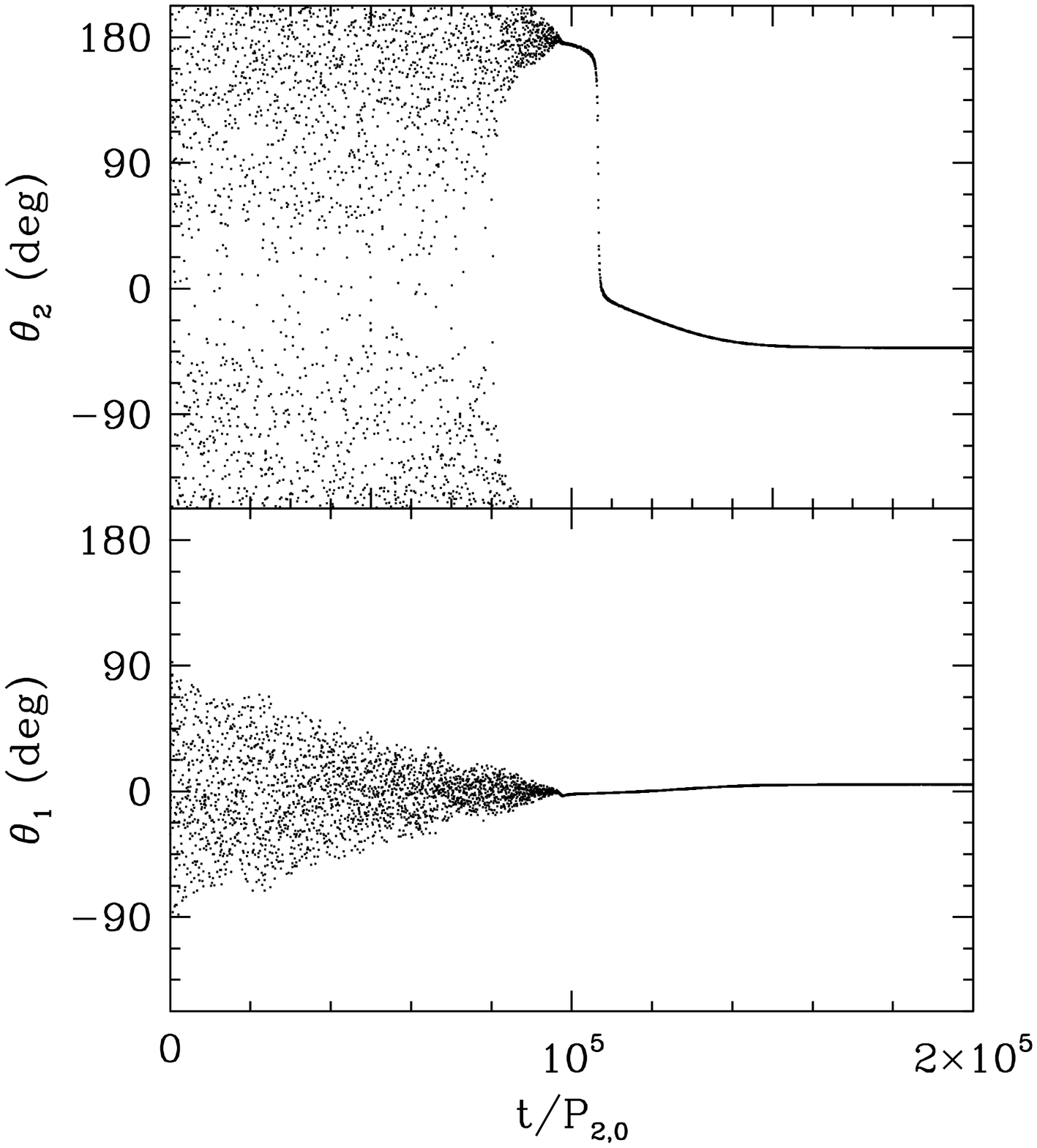}
\caption{
Evolution of the semimajor axes $a_1$ and $a_2$, eccentricities $e_1$
and $e_2$, departure of the period ratio $P_2/P_1$ from $2$, and 2:1
eccentricity-type resonance variables $\theta_1 = \lambda_1 - 2
\lambda_2 + \varpi_1$ and  $\theta_2 = \lambda_1 - 2 \lambda_2 +
\varpi_2$, for a convergent migration calculation with the stellar
mass $M_\ast = 1 M_\sun$ and planetary masses $M_1 = M_2 = 10 M_\oplus$.
The outer planet is forced to migrate inward with 
${\dot a}_2/a_2 = -2\times 10^{-6}/P_2$ and eccentricity damping
factor $K_e = 100$.
The semimajor axes and time are in units of the initial orbital
semimajor axis, $a_{2,0}$, and period, $P_{2,0}$, of the outer planet,
respectively.
}
\label{fig:capture}
\end{figure}

We have performed direct numerical orbit integrations using the
symplectic integrator SyMBA modified to include forced migration and
eccentricity damping \citep{Lee2002,Lee2004}.
Figure \ref{fig:capture} shows a convergent migration calculation
with $M_\ast = 1 M_\sun$ and $M_1 = M_2 = 10 M_\oplus$.
The planets are initially far from the 2:1 mean-motion
commensurability, and the outer planet is forced to migrate inward
with ${\dot a}_2/a_2 \propto P_2^{-1}$ and $K_e = 100$.
The pair is captured into 2:1 resonance with both $\theta_1$ and
$\theta_2$ librating.
The centers of libration change from $\theta_1 = 0^\circ$ and
$\theta_2 = 180^\circ$ at small eccentricities to close to
$\theta_1 = \theta_2 = 0^\circ$ at large eccentricities (with the
offsets from $0^\circ$ due to the forced migration and eccentricity
damping).
The eccentricities reach equilibrium values that are too large ($e_1
\approx 0.2$), and $P_2/P_1$ departs from 2 by less than $0.001$ at
the end.
This result is representative of calculations with $K_e \sim 100$ for
both 2:1 and 3:2.
Hence the Kepler near-resonance pairs are {\it not} the result of
eccentricity damping within the expected range during disk-induced
classic type I migration.

\section{TIDAL DAMPING OF ECCENTRICITY}

It has been proposed that the subsequent damping of orbital
eccentricities by tidal dissipation in the planets may reduce the
eccentricities to sufficiently small values to explain the observed
departures from exact commensurabilities
\citep{Lithwick2012a,Batygin2013}.
Tidal dissipation in the planet damps the orbital eccentricity on
timescale
\begin{equation}
\tau_e = {e \over {\dot e}}
= {1 \over 21\pi} {Q \over k_2} {M_p \over M_\ast} \left(a \over
R_p\right)^5 P ,
\label{eq:taue}
\end{equation}
while conserving the total angular momentum of the system, where
$R_p$, $Q$, and $k_2$ are the radius, tidal dissipation function, and
Love number of the planet of mass $M_p$.
The above equation assumes that the planet is synchronously rotating
and that $Q$ is constant as a function of the tidal frequencies.
Since ${\dot e}/e$ is independent of $e$ and $a$ changes only slightly
for small $e$, one would expect $e$ to decay exponentially.
However, as we now demonstrate, the eccentricities do {\it not} decay
exponentially for near-resonance pairs due to interactions between the
planets: they decay slowly according to a shallow power law.

\begin{figure}[btp]
\epsscale{0.8}
\plotone{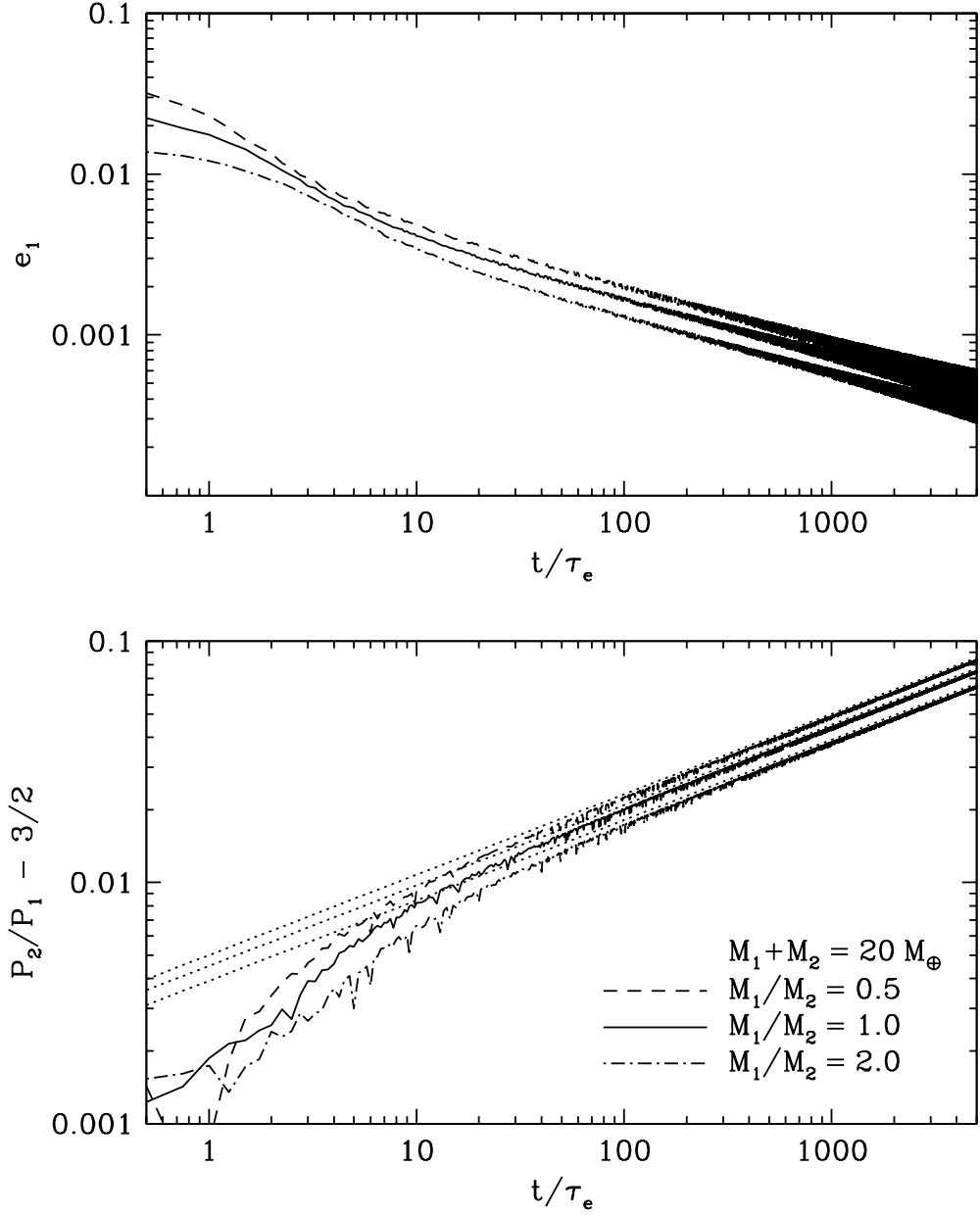}
\caption{
Evolution of the departure of the period ratio $P_2/P_1$ from $3/2$
and eccentricity of the inner planet $e_1$ for numerical calculations
where $e_1$ is damped on a constant timescale $\tau_e$ to simulate
tidal dissipation in the inner planet.
The planet pair is initially in the 3:2 resonance, with $M_\ast = 1
M_\sun$, $M_1 + M_2 = 20 M_\oplus$, and $M_1/M_2 = 0.5$ (dashed
lines), $1.0$ (solid lines), and $2.0$ (dot-dashed lines).
The dotted lines in the lower panel show Equations
(\ref{eq:pratiopowerlaw}) and (\ref{eq:Dj}) from the analytic theory.
}
\label{fig:e1dP32}
\end{figure}

\begin{figure}[btp]
\epsscale{0.8}
\plotone{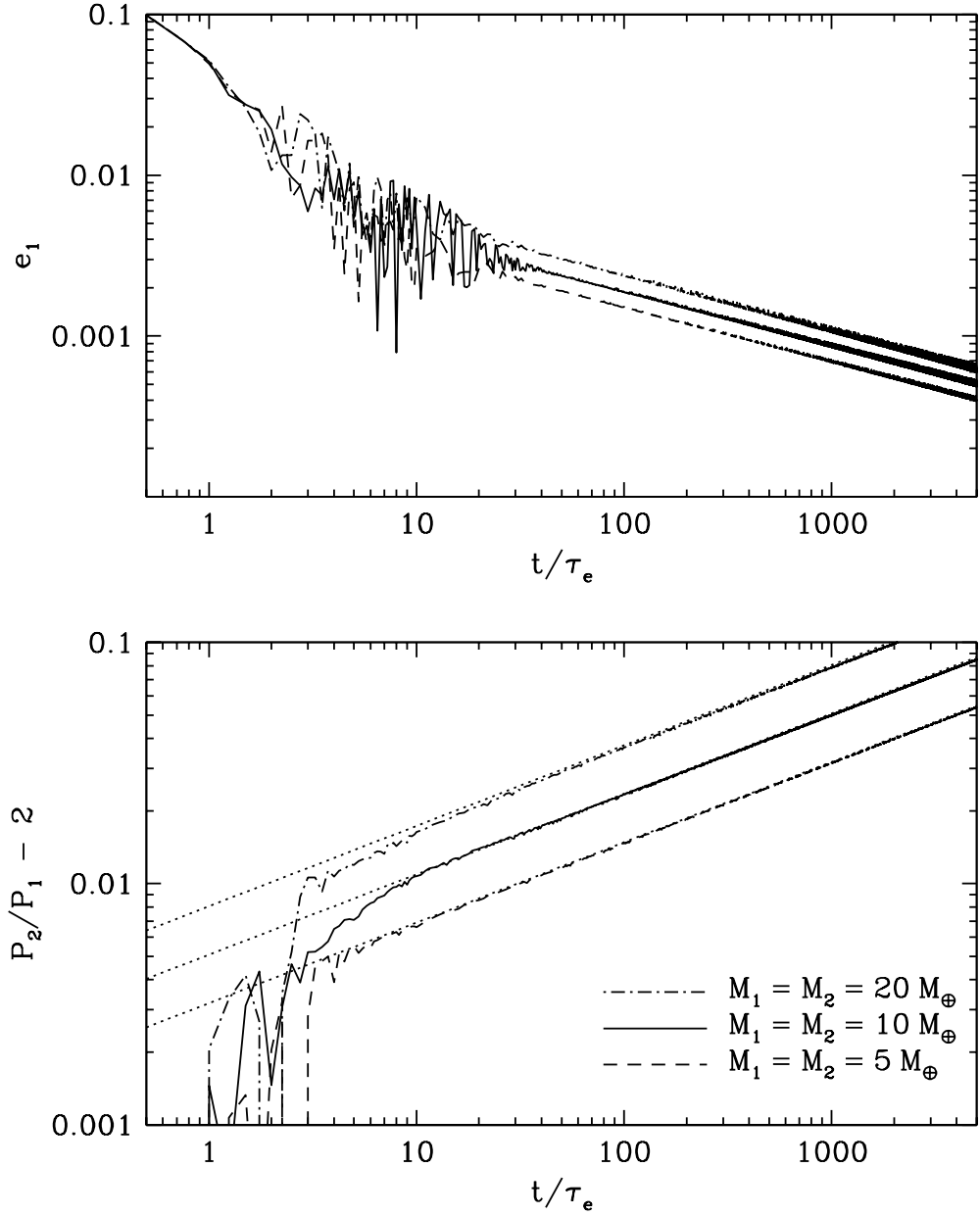}
\caption{
Same as Figure \ref{fig:e1dP32}, but for the 2:1 resonance with $M_1
= M_2 = 5 M_\oplus$ (dashed lines), $10 M_\oplus$ (solid lines), and
$20 M_\oplus$ (dot-dashed lines).
}
\label{fig:e1dP21}
\end{figure}

Using migration calculations with eccentricity damping similar to that
shown in Figure \ref{fig:capture}, we have assembled several 2:1 and
3:2 configurations with $M_\ast = 1 M_\sun$ and different $M_1$ and
$M_2$.
These configurations were then evolved in calculations where the
eccentricity of the inner planet is damped on a constant timescale
$\tau_e$ (while the semimajor axis of the inner planet is adjusted at
the same time to conserve orbital angular momentum) to simulate tidal
dissipation in the inner planet.
We ignore tidal dissipation in the outer planet as the rate is a steep
function of semimajor axis (Equation (\ref{eq:taue})).
The results are plotted in Figures \ref{fig:e1dP32} and
\ref{fig:e1dP21}.
In Figure \ref{fig:e1dP32} for 3:2, the dashed, solid, and dot-dashed
lines are for $M_1 + M_2 = 20 M_\oplus$ and $M_1/M_2 = 0.5$, $1.0$,
and $2.0$, respectively.
In Figure \ref{fig:e1dP21} for 2:1, the dashed, solid, and dot-dashed
lines are for $M_1 = M_2 = 5$, $10$, and $20 M_\oplus$, respectively.
The figures show that
\begin{equation}
e_1 \propto (t/\tau_e)^{-1/3} ,
\end{equation}
and
\begin{equation}
{P_2 \over P_1} - {j \over (j-1)} = (D_j t/\tau_e)^{1/3} ,
\label{eq:pratiopowerlaw}
\end{equation}
after an initial transient period of a few $\tau_e$.
We have checked that the relationships in Equations (\ref{eq:e2e1})
and (\ref{eq:pratioe1}) are satisfied during the power-law decay.
\cite{Lithwick2012a}, \cite{Batygin2013}, and \cite{Delisle2012} have
explained this power-law behavior (see also \citealt{Papaloizou2010})
and derived analytically $D_j$ as a function of $M_1/M_\ast$ and
$M_2/M_1$ for the $j$:$(j-1)$ resonance:
\begin{equation}
D_j = {9 j^2 \over (j - 1)^3} \left(M_1 \over M_\ast\right)^2 \beta
      (1 + \beta) C_1^2 ,
\label{eq:Dj}
\end{equation}
where
\begin{equation}
\beta = {M_2 \over M_1} \alpha^{-1/2}
= {M_2 \over M_1} \left(j \over j - 1\right)^{1/3} .
\end{equation}
The dotted lines in Figures \ref{fig:e1dP32} and \ref{fig:e1dP21} show
the analytic result, which is in excellent agreement with the
numerical results after the initial transient period.

We have so far assumed that tidal dissipation in the inner planet
conserves the orbital angular momentum.
Strictly speaking, tidal dissipation conserves the total angular
momentum, which includes the spin angular momentum.
There is a small change in $D_j$ if we account for the tidal evolution
of the planet's spin.
We have also ignored the tidal dissipation in the outer planet, which
adds to $D_j/\tau_e$ but does not change the $1/3$ power-law behavior
\citep{Lithwick2012a,Batygin2013}.

An important consequence of this slow power-law behavior is that many
$\tau_e$ must elapse to produce a departure of $P_2/P_1$ of a few
percent from exact commensurability.
For example, $P_2/P_1 - j/(j-1) \approx 0.03$ requires $t \ga 50
\tau_e$ (see Figs.~\ref{fig:e1dP32} and \ref{fig:e1dP21}).
For $P = 10\,$days, $M_\ast = 1 M_\sun$, $M_p = 10 M_\oplus$, $R_p = 3
R_\oplus$,
\begin{equation}
\tau_e = 2.26 \times 10^6 (Q/k_2)\,{\rm yr} .
\label{eq:taue2}
\end{equation}
Whether a near-resonance pair can reach its $P_2/P_1 - j/(j-1)$ within
the age of its host star ($\sim$ a few Gyr) depends critically on
$Q/k_2$ of the inner planet, which is very different for rocky and
giant planets, as we now review.

\subsection{Known Constraints on $\mbox{\boldmath $Q/k_2$}$ of Planets}
\label{sec:Qk2}

The tidal $Q/k_2$ of Solar System planets have been measured or
constrained by the tidal evolution of their satellites.
The known value and limit on $Q/k_2$ for rocky planets are $Q/k_2 =
40$ for Earth \citep{Murray1999} and $470 < Q/k_2 < 1000$ for Mars
\citep{Shor1975,Sinclair1978,Duxbury1982}.
The known limits on $Q/k_2$ for giant planets are:
$1.6 \times 10^5 < Q/k_2 < 5.3 \times 10^6$ for Jupiter
\citep{Yoder1981},
$5.4 \times 10^4 < Q/k_2 < 2.9 \times 10^5$ for Saturn
\citep{Peale1999,Meyer2008},
$1.1 \times 10^5 < Q/k_2 < 3.8 \times 10^5$ for Uranus
\citep{Tittemore1990}, and
$2.2 \times 10^4 < Q/k_2 < 8.8 \times 10^4$ for Neptune
\citep{Banfield1992,Zhang2008}.
For extrasolar giant planets, $6.7 \times 10^4 < Q/k_2 < 6.7 \times
10^8$ has been derived from the existence of some close-in planets
with non-zero orbital eccentricities \citep{Matsumura2008}.
So the {\it lowest} bound on $Q/k_2$ is $40$ for rocky planets from
Earth and $2.2 \times 10^4$ for giant planets from Neptune.

If we substitute these lowest bounds into the estimate in Equation
(\ref{eq:taue2}), we find $\tau_e \ga 9 \times 10^7\,{\rm yr}$ for
rocky planets and $\tau_e \ga 5 \times 10^{10}\,{\rm yr}$ for giant
planets, which indicate that a near-resonance pair might be able to
reach its $P_2/P_1 - j/(j-1)$ over the age of its host star, if the
inner planet is rocky (and not if the inner planet is a giant).

\section{COMPARISON WITH OBSERVATIONS}

\begin{figure}[btp]
\epsscale{0.7}
\plotone{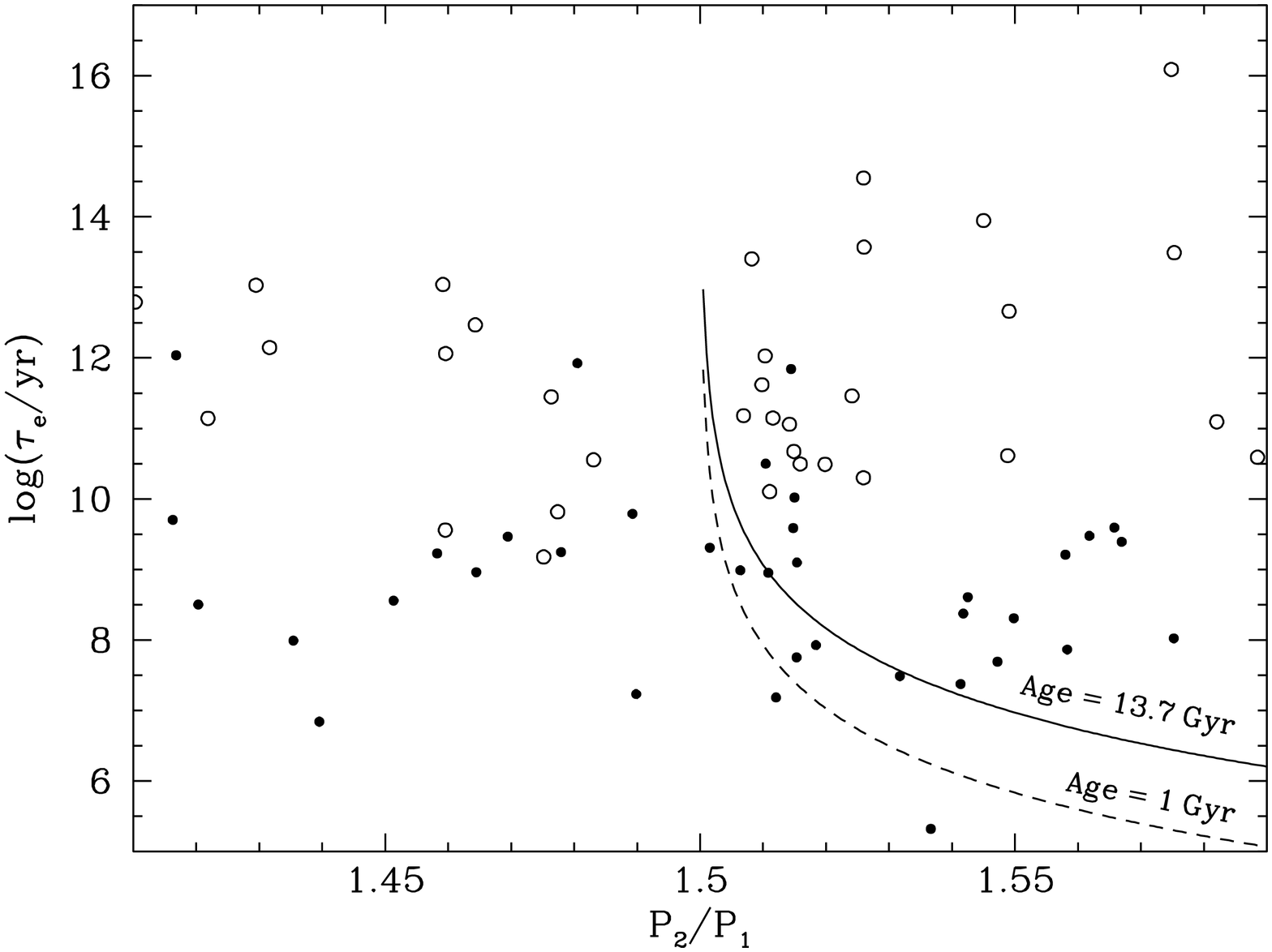}
\medskip
\plotone{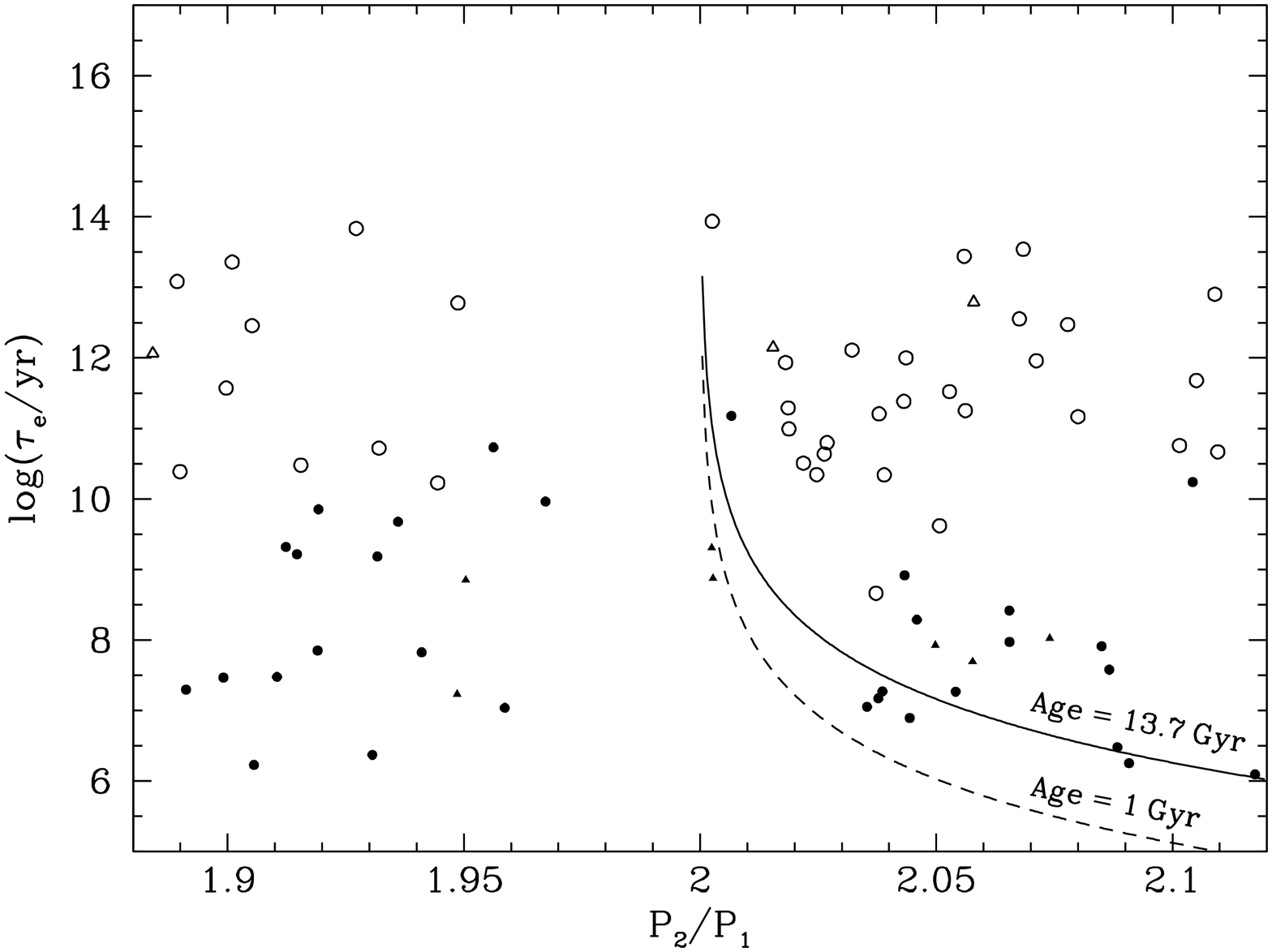}
\caption{
Tidal eccentricity damping timescale $\tau_e$ (Equation
(\ref{eq:taue})) of the inner planet for the Kepler candidate pairs
near the 2:1 (lower panel) and 3:2 (upper panel) resonances.
Circles and triangles are adjacent and non-adjacent pairs,
respectively, and filled and open symbols are pairs with $R_1 < 2
R_\oplus$ and $R_1 > 2 R_\oplus$, respectively.
We adopt $Q/k_2 = 10^5$ and $M_p = M_\oplus (R_p/R_\oplus)^{2.06}$
for planets with $R_p > 2 R_\oplus$, and $Q/k_2 = 100$ and $M_p =
M_\oplus (R_p/R_\oplus)^{3.7}$ for planets with $R_p < 2 R_\oplus$.
The dashed and solid lines show $\tau_e$ as a
function of $P_2/P_1$ according to Equations (\ref{eq:pratiopowerlaw})
and (\ref{eq:Dj}) for $t = 1$ and $13.7\,$Gyr, respectively, if we
have two $10 M_\oplus$ planets orbiting a solar-mass star.
}
\label{fig:taue}
\end{figure}

Figure \ref{fig:taue} shows the tidal eccentricity damping timescale
$\tau _e$ (Equation (\ref{eq:taue})) of the inner planet for the Kepler
candidate pairs near the 2:1 and 3:2 resonances.
Circles are adjacent pairs, and triangles are non-adjacent pairs.
Filled and open symbols are pairs with the radius of the inner planet
$R_1 < 2 R_\oplus$ and $R_1 > 2 R_\oplus$, respectively.
For the ``giant'' planets with $R_p > 2 R_\oplus$, we adopt $Q/k_2 =
10^5$ and mass from the mass-radius relationship $M_p = M_\oplus
(R_p/R_\oplus)^{2.06}$ of \cite{Lissauer2011}, which is consistent
with Earth to Saturn in the Solar System (other proposed mass-radius
relationships, such as those of \citealt{Wu2013} and
\citealt{Weiss2013}, would give a similar plot).
For the ``rocky'' planets with $R_p < 2 R_\oplus$, we adopt $Q/k_2 =
100$ and mass from the mass-radius relationship $M_p = M_\oplus
(R_p/R_\oplus)^{3.7}$ of \cite{Valencia2006}.
The dashed and solid lines in Figure \ref{fig:taue} show $\tau_e$ as a
function of $P_2/P_1$ according to equations (\ref{eq:pratiopowerlaw})
and (\ref{eq:Dj}) for $t = 1\,$Gyr and $13.7\,$Gyr, respectively, if
we have two $10 M_\oplus$ planets orbiting a solar-mass star  (i.e.,
$M_1/M_\ast = 3 \times 10^{-5}$ and $M_1 = M_2$).
These lines indicate where such a resonant pair would be in $P_2/P_1$,
if it started near exact commensurability, its age is $t = 1\,$Gyr or
$13.7\,$Gyr, and the tidal eccentricity damping timescale of the inner
planet is $\tau_e$.
There are some filled symbols below the solid lines for the age of the
Universe, hinting that some of the near-resonance pairs with $R_1 < 2
R_\oplus$ can potentially reach their current locations by tidal
eccentricity damping in less than the age of the host star.
Most of the pairs with $R_1 > 2 R_\oplus$ (open symbols) are well
above the lines, indicating that they are unlikely to reach their
current locations by tidal eccentricity damping.
However, the comparison in Figure \ref{fig:taue} is not exact, as the
theoretical curves are for a specific combination of stellar and
planetary masses, and the observed Kepler pairs are plotted for
assumed $Q/k_2$.

Turning the inference around, we may assume that the current
architecture was established by tides, and thereby infer constraints
on $Q/k_2$.
We can plot (Equations (\ref{eq:taue}) and (\ref{eq:pratiopowerlaw}))
\begin{equation}
{Q \over k_2} = {21\pi \over P_1} {M_\ast \over M_1} \left(R_1
\over a\right)^{5} \left[P_2/P_1 - j/(j-1)\right]^{-3} D_j t
\label{eq:qk2max}
\end{equation}
versus $P_2/P_1$ near the $j$:$(j-1)$ resonance with $t = 13.7\,$Gyr.
This is the {\it maximum} $Q/k_2$ or minimum tidal dissipation
efficiency that the inner planet must have if the pair is to evolve to
its current $P_2/P_1$ in less than the age of the Universe.
The actual $Q/k_2$ is smaller by the ratio of the age of the planetary
system to the age of the Universe.
In Figure \ref{fig:qk2max}, the points show this maximum $Q/k_2$ for
the inner planet of the observed Kepler pairs.
The four panels show the $R_1 < 2 R_\oplus$ and $R_1 > 2 R_\oplus$
cases for 3:2 and 2:1.
The lines with arrows pointing upward are the known lowest bound on
$Q/k_2$ for planets ($2.2 \times 10^4$ for giant planets and $40$ for
rocky planets; see \S \ref{sec:Qk2}).
Figure \ref{fig:qk2max} clearly shows that some pairs with $R_1 < 2
R_\oplus$ can reach where they are by tidal eccentricity damping, if
planets with $R_p < 2 R_\oplus$ are rocky with $Q/k_2 \ge 40$.
They include KOI~500.03/04, KOI~500.04/01, KOI~730.02/01,
KOI~961.01/03, and KOI~2038.01/02 just outside 3:2, and KOI~720.04/01,
KOI~904.01/04, KOI~952.04/01, KOI~1161.03/01, and KOI~1824.02/01 just
outside 2:1, which are $\sim 20$--$33\%$ of all pairs with $R_1 < 2
R_\oplus$ in the period ratio ranges shown in Figure \ref{fig:qk2max}.
However, none of the pairs with $R_1 > 2 R_\oplus$ can reach their
current $P_2/P_1$, if planets with $R_p > 2 R_\oplus$ are giants with
$Q/k_2 \ge 2.2 \times 10^4$.
Furthermore, there are clumps of $R_1 > 2 R_\oplus$ pairs just outside
2:1 and 3:2 that are more than an order of magnitude below $Q/k_2= 2.2
\times 10^4$.

In Figure \ref{fig:qk2max} we only plot the adjacent pairs, as the
tidal evolution of the non-adjacent pairs can be significantly
affected by secular or even resonant interactions with the intervening
planet(s).
Adjacent pairs with $R_1 > 2 R_\oplus$ could evolve significantly
faster, if the outer planet has $R_2 < 2 R_\oplus$ and much lower
$Q/k_2$ to overcome the steep dependence of the tidal eccentricity
damping rate on orbital semimajor axis (Equation (\ref{eq:taue})).
However, for the observed Kepler pairs with $R_1 > 2 R_\oplus$ in the
upper panels of Figure \ref{fig:qk2max}, only three with $P_2/P_1$
between $1.5$ and $1.53$ and two with $P_2/P_1$ between $2$ and $2.06$
have an outer planet with $R_2 < 2 R_\oplus$.
So our conclusion that most of the pairs with $R_1 > 2 R_\oplus$
cannot reach their current $P_2/P_1$ by tidal eccentricity damping, if
planets with $R_p > 2 R_\oplus$ are giants with $Q/k_2 \ge 2.2 \times
10^4$, is robust.

In the above analysis, we follow the Kepler team in using $R_p = 2
R_\oplus$ as the boundary between super-Earths and Neptunes.
Some pairs are moved from the upper panels of Figure \ref{fig:qk2max}
to the lower panels if we increase this boundary to $R_p = 2.25
R_\oplus$ (and vice versa if we decrease this boundary to $R_p = 1.75
R_\oplus$), but the overall patterns (i.e., some pairs in the lower
panels have maximum $Q/k_2$ of the inner planet above $40$ and all
pairs in the upper panels have maximum $Q/k_2$ of the inner planet
below $2.2 \times 10^4$) remain the same.

\section{SUMMARY AND DISCUSSION}

We have shown that some of the Kepler near-resonance pairs with
$R_1 < 2 R_\oplus$ may be able to move to their current near-resonance
locations by tidal damping of eccentricity if they are rocky with
$Q/k_2 \sim 100$, but that all known pairs with $R_1 > 2 R_\oplus$ are
unable to move to their current near-resonance locations by the same
mechanism if they are giants with $Q/k_2 \ga 2 \times 10^4$.

\begin{figure}[btp]
\epsscale{1.0}       
\plotone{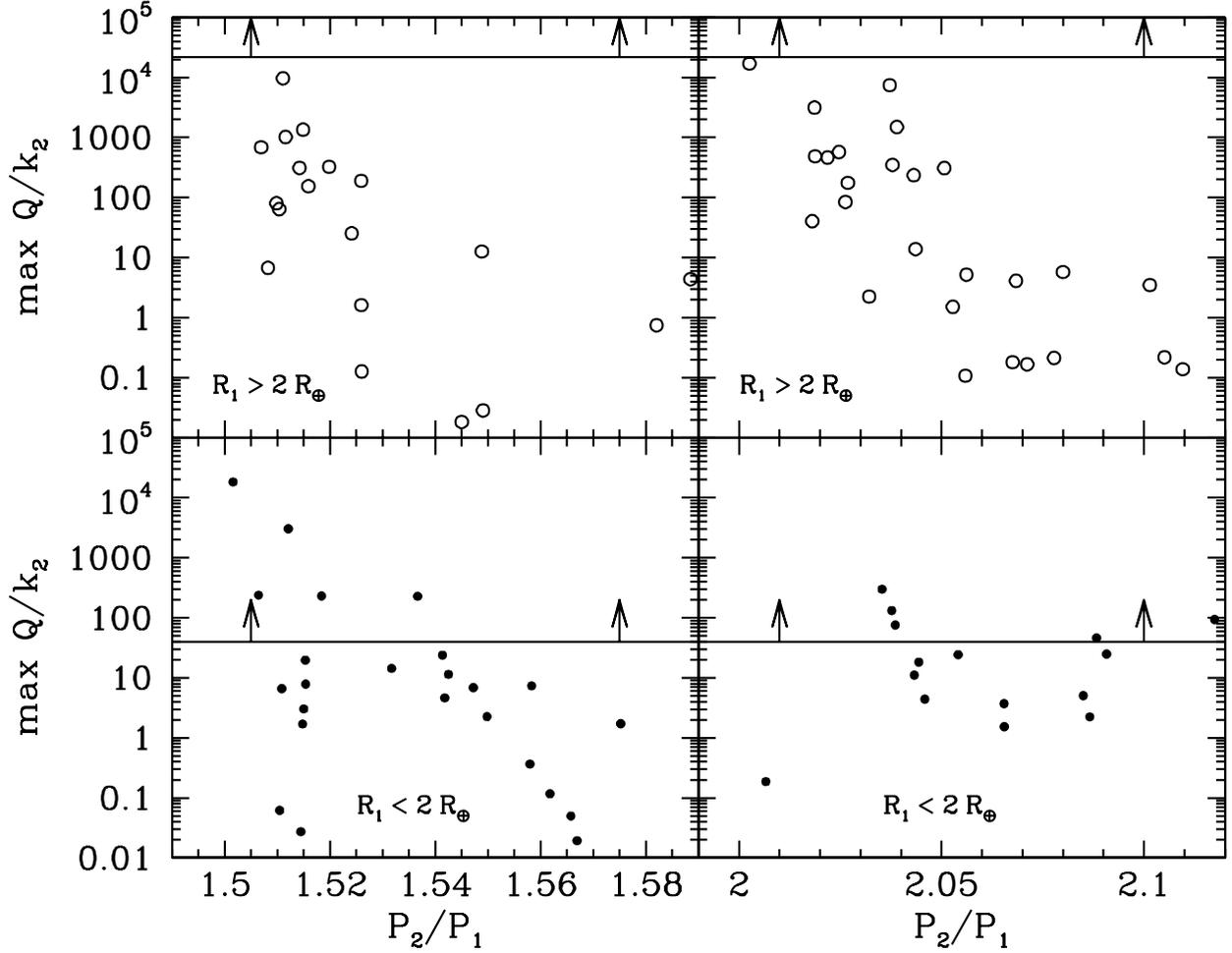}
\caption{
Maximum $Q/k_2$ (Equation (\ref{eq:qk2max})) that the inner planet
must have if the Kepler near-resonance pairs are to evolve to their
current $P_2/P_1$ in less than $13.7\,$Gyr.
The pairs with $R_1 < 2 R_\oplus$ (lower panels) and $R_1 > 2
R_\oplus$ (upper panels) are shown in the left and right panels for
the 3:2 and 2:1 resonances, respectively.
We adopt $M_p = M_\oplus (R_p/R_\oplus)^{2.06}$
for planets with $R_p > 2 R_\oplus$, and $Q/k_2 = 100$ and $M_p =
M_\oplus (R_p/R_\oplus)^{3.7}$ for planets with $R_p < 2 R_\oplus$.
The lines with arrows pointing upward are the known lowest bound on
$Q/k_2$: $2.2 \times 10^4$ for giant planets and $40$ for rocky
planets.
}
\label{fig:qk2max}
\end{figure}

\begin{figure}[btp]
\epsscale{1.0}       
\plotone{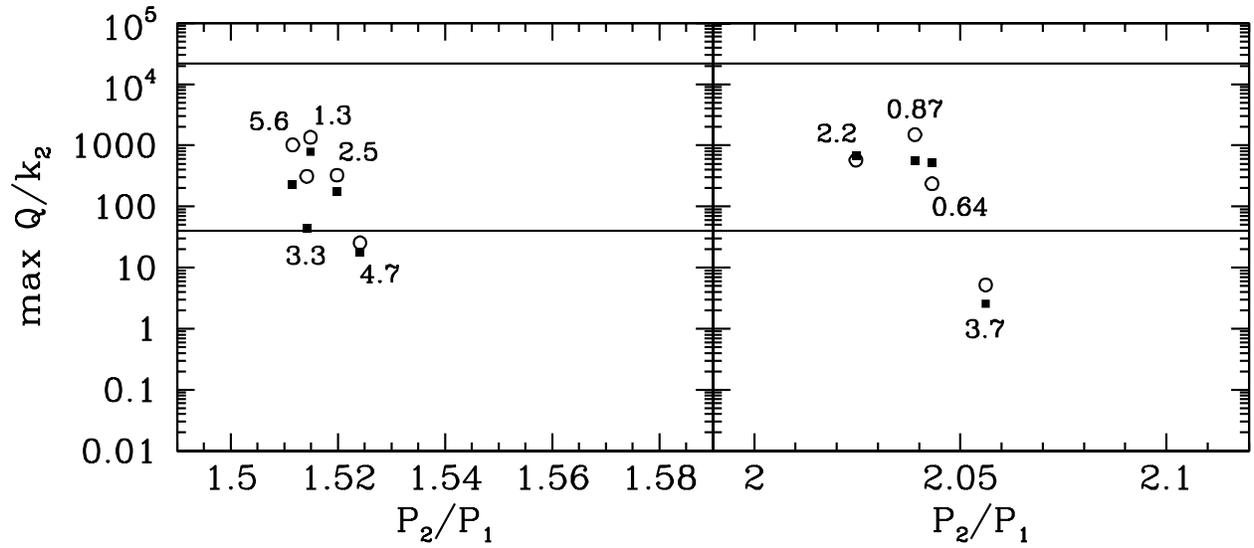}
\caption{
Same as Figure \ref{fig:qk2max}, but for the pairs with planetary
masses obtained from transit timing variations (TTV) by \cite{Wu2013}.
All pairs in this sample  have  $R_1 > 2 R_\oplus$.
Each pair shifts vertically from the maximum $Q/k_2$ determined using
assumed mass-radius relationships (open circle; same as in the upper
panels of Figure \ref{fig:qk2max}) to that using the actual masses
from TTV (filled square).
The numerical value next to each pair indicates the mean density (in
g$\,$cm$^{-3}$) of the inner planet from TTV.
}
\label{fig:qk2max3}
\end{figure}

What are the alternatives?
One possibility is that some of the $R_p > 2 R_\oplus$ planets are in
fact rocky with low $Q/k_2$.
We have checked that increasing the boundary between super-Earths and
Neptunes to, say, $R_p = 2.25 R_\oplus$ does not change our
conclusions.
This possibility can also be checked by measuring or constraining the
masses $M_p$, and hence the mean densities $\rho_p$, of the planets
using transit timing variations (TTV) or radial velocity data.
\cite{Wu2013} have determined $M_p$ and $\rho_p$ for $16$ pairs of
Kepler planets from TTV (see also \citealt{Lithwick2012b,Xie2012}).
Nine of these pairs are in the period ratio ranges plotted in Figure
\ref{fig:qk2max}, all with $R_1 > 2 R_\oplus$.
Figure \ref{fig:qk2max3} shows the maximum $Q/k_2$ of the inner planet
for these nine pairs according to Equation (\ref{eq:qk2max}), with the
open circles using assumed mass-radius relationships (same as in the
upper panels of Figure \ref{fig:qk2max}) and the filled squares using
the actual masses determined from TTV.
The maximum $Q/k_2$ values shift when actual masses are used, but
they remain more than an order of magnitude below the lowest bound
$Q/k_2 = 2.2 \times 10^4$ for giant planets, and mostly above the
lowest bound $Q/k_2 = 40$ for rocky planets.
Although the inner planets of two pairs (KOI~1336.01/02 and
KOI~168.03/01) have $\rho_p \ge 4.7{\rm\,g}{\rm\,cm}^{-3}$ that
clearly exceed the bulk densities of uncompressed rocks, the inner
planets of three pairs (KOI~841.01/02, KOI~244.02/01, and
KOI~248.01/02) have $\rho_p = 0.64$--$1.3{\rm\,g}{\rm\,cm}^{-3}$,
which are more consistent with giant planets.
Although \cite{Wu2013} have suggested that the low densities of some
planets with $R_p \le 3 R_\oplus$ (such as KOI~244.02 and KOI~248.01)
could be due to an extended envelope of hydrogen and helium $\la 1\%$
in mass on a rocky planet, KOI~841.01 (which has $R_p = 5.44 R_\oplus$
and $\rho_p = 0.64{\rm\,g}{\rm\,cm}^{-3}$) is most likely a giant
planet.

Another relevant case is the KOI~142 system, which has one transiting
planetary candidate (KOI~142.01) that shows large TTV as well as
detectable transit duration variations (TDV).
\cite{Nesvorny2013} have determined from the TTV and TDV that
KOI~142.01 has $R_p = 4.23_{-0.39}^{+0.30} R_\oplus$ and $\rho_p =
0.48_{-0.46}^{+0.54}{\rm\,g}{\rm\,cm}^{-3}$ and that it is just wide
of the 2:1 resonance ($P_2/P_1 = 2.03$) with an outer planet with $M_2
= 216 M_\oplus$.
Since there is also an estimated stellar age ($\approx 2.45\,$Gyr), we
can determine from Equation (\ref{eq:qk2max}) the actual (instead of
maximum) $Q/k_2$ needed for KOI~142.01: 3400, which is lower than the
lowest bound for giant planets by a factor of $6.5$.
However, this constraint on $Q/k_2$ of KOI~142.01 is unlikely to be
valid, because the relatively large orbital eccentricities of both
planets (mean $e_1 = 0.064$ and $e_2 = 0.055$) are not consistent with
the tidal eccentricity damping scenario \citep{Nesvorny2013}.

So another mechanism is needed for some of the pairs with $R_1 > 2
R_\oplus$.
\cite{Petrovich2013} have considered the possibility of {\it in situ}
formation of planets near first-order mean-motion resonances in a
simple dynamical model without orbital migration or dissipation.
For the effective viscosity and dimensionless scale height, $H/a$,
typically assumed for protoplanetary disks, most of the Kepler
candidates are sufficiently small that they are unable to open gaps in
their natal protoplanetary disks and should undergo type I migration.
\cite{Rein2012} has suggested migration in a turbulent disk, which has
both smooth and stochastic components, as a way to produce
near-resonance pairs.
The departure from exact commensurability can be used to constrain the
relative strength of smooth and stochastic migration.
However, \cite{Rein2012} has assumed that the smooth migration is
always inward, which is only true for classic type I (and type II)
migration.
Alternatively, \cite{Baruteau2013} have shown that planet-disk
interactions for partial gap-opening planets may provide sufficient
energy dissipation and eccentricity damping and lead to near-resonance
pairs.
However, effective viscosity and/or dimensionless scale height smaller
than typical values is required for most of the Kepler candidates to
open partial gaps.

Finally, we note that recent analysis of the corotation and horseshoe
torques (plus the differential Lindblad torque) have shown that the
coefficient $C_a$ for the migration rate in Equation (\ref{eq:dadt})
is a function of the local surface density gradient $d\ln\Sigma/d\ln
a$, temperature gradient $d\ln T/d\ln a$, viscous saturation parameter
$p_\nu$, and thermal saturation parameter $p_\chi$ (e.g.,
\citealt{Paardekooper2010,Paardekooper2011}).
In certain disk models (e.g., \citealt{Garaud2007}), it is possible
for type I migration to be outward in the viscously heated regions of
the disk and in the region inside the magnetospheric truncation
radius, if the corotation and horseshoe torques are not saturated
\citep{Paardekooper2011,Kretke2012}.
There are also locations in the disk where the total torque vanishes
and the migration is stalled.
This more complex migration behavior means that it is possible for a
pair of planets to undergo both convergent and divergent migration, as
the disk accretion rate decreases with time and the disk depletes.
Whether the breaking of resonances by divergent migration could result
in an excess of planet pairs just outside the first-order resonances
will require further investigation.
In this paper, however, we have shown that the later evolution due to
tides is not enough to explain the structures near resonances.

\acknowledgments
We thank the referee for helpful comments on the manuscript.
M.H.L.\ was supported by Hong Kong RGC grant HKU 7034/09P.
D.F.\ was supported by NASA through Hubble Fellowship grant
HF-51272.01-A awarded by STScI.
D.N.C.L.\ was supported by NASA (NNX08AM84G), NSF (AST-0908807), and a
University of California Lab Fee grant.


{}

\clearpage

\clearpage

\clearpage

\clearpage

\clearpage

\clearpage

\begin{thebibliography}{}

\bibitem[Artymowicz(1993)]{Artymowicz1993}
Artymowicz, P. 1993, \apj, 419, 166

\bibitem[Banfield \& Murray(1992)]{Banfield1992}
Banfield, D., \& Murray, N. 1992, Icarus, 99, 390

\bibitem[Baruteau \& Papaloizou(2013)]{Baruteau2013}
Baruteau, C., \& Papaloizou, J. C. B. 2013, \apj, submitted
(arXiv:1301.0779)

\bibitem[Batalha et al.(2013)]{Batalha2013}
Batalha, N. M.., et al. 2013, \apjs, 204,24

\bibitem[Batygin \& Morbidelli(2013)]{Batygin2013}
Batygin, K., \& Morbidelli, A. 2013, \aj, 145, 1

\bibitem[Bryden et al.(2000)]{Bryden2000}
Bryden, G., R\'o\.zyczka, M., Lin, D. N. C., \& Bodenheimer, P. 2000,
\apj, 540, 1091

\bibitem[Burke et al.(2013)]{Burke2013}
Burke, C. J., et al. 2013, AAS Meeting, 221, 216.02

\bibitem[Delisle et al.(2012)]{Delisle2012}
Delisle, J.-B., Laskar, J., Correia, A. C. M., \& Bou\'e, G. 2012,
\aap, 546, 71

\bibitem[Duxbury \& Callahan(1982)]{Duxbury1982}
Duxbury, T. C., \& Callahan, J. D. 1982, Lunar Planet. Sci. Conf., 13, 191

\bibitem[Fabrycky et al.(2012)]{Fabrycky2012}
Fabrycky, D. C., et al. 2012, \apj, submitted (arXiv:1202.6328)

\bibitem[Garaud \& Lin(2007)]{Garaud2007}
Garaud, P., \& Lin, D. N. C. 2007, \apj, 654, 606

\bibitem[Goldreich \& Tremaine(1980)]{Goldreich1980}
Goldreich, P., \& Tremaine, S. 1980, \apj, 241, 425

\bibitem[Hut(1981)]{Hut1981}
Hut, P. 1981, \aap, 99, 126

\bibitem[Kley(2000)]{Kley2000}
Kley, W. 2000, \mnras, 313, L47

\bibitem[Kretke \& Lin(2012)]{Kretke2012}
Kretke, K. A., \& Lin, D. N. C. 2012, \apj, 755, 74

\bibitem[Laughlin \& Chambers(2001)]{Laughlin2001}
Laughlin, G., \& Chambers, J. E. 2001, \apjl, 551, L109

\bibitem[Laughlin et al.(2005)]{Laughlin2005}
Laughlin, G., Butler, R. P., Fischer, D. A., Marcy, G. W., Vogt,
S. S., \& Wolf, A. S. 2005, \apj, 622, 1182

\bibitem[Lee(2004)]{Lee2004}
Lee, M. H. 2004, \apj, 611, 517

\bibitem[Lee \& Peale(2002)]{Lee2002}
Lee, M. H., \& Peale, S. J. 2002, \apj, 567, 596

\bibitem[Lee et al.(2006)]{Lee2006}
Lee, M. H., Butler, R. P., Fischer, D. A., Marcy, G. W., \& Vogt S. S.
2006, \apj, 641, 1178

\bibitem[Lin \& Papaloizou(1986)]{Lin1986}
Lin, D. N. C., \& Papaloizou, J. 1986, \apj, 309, 846

\bibitem[Lissauer et al.(2011)]{Lissauer2011}
Lissauer, J. J., et al. 2011, \apjs, 197, 8

\bibitem[Lissauer et al.(2012)]{Lissauer2012}
Lissauer, J. J., et al. 2012, \apj, 750, 112

\bibitem[Lithwick \& Wu(2012)]{Lithwick2012a}
Lithwick, Y., \& Wu, Y. 2012, \apj, 756, L11

\bibitem[Lithwick et al.(2012)]{Lithwick2012b}
Lithwick, Y., Xie, J., \& Wu, Y. 2012, \apj, 761, 122

\bibitem[Matsumura et al.(2008)]{Matsumura2008}
Matsumura, S., Genya, T., \& Rasio, F. A. 2008, \apj, 686, L29

\bibitem[Meyer \& Wisdom(2008)]{Meyer2008}
Meyer, J., \& Wisdom, J. 2008, Icarus, 193, 178

\bibitem[Murray \& Dermott(1999)]{Murray1999}
Murray, C. D., \& Dermott, S. F. 1999, Solar System Dynamics
(Cambridge: Cambridge Univ. Press)

\bibitem[Nesvorn\'y et al.(2013)]{Nesvorny2013}
Nesvorn\'y, D., Kipping, D., Terrell, D., Hartman, J., Bakos, G. A.,
\& Buchhave, L. A. 2013, preprint (arXiv:1304.4283)

\bibitem[Novak et al.(2003)]{Novak2003}
Novak, G. S., Lai, D., \& Lin, D. N. C. 2003, in ASP Conf.\ Ser.\ 294,
Scientific Frontiers in Research on Extrasolar Planets, ed. D. Deming
\& S. Seager (San Francisco: ASP), 177

\bibitem[Paardekooper et al.(2010)]{Paardekooper2010}
Paardekooper, S.-J., Baruteau, C., Crida, A., \& Kley, W. 2010,
\mnras, 401, 1950

\bibitem[Paardekooper et al.(2011)]{Paardekooper2011}
Paardekooper, S.-J., Baruteau, C., \& Kley, W. 2011, \mnras, 410, 293

\bibitem[Papaloizou \& Terquem(2010)]{Papaloizou2010}
Papaloizou, J. C. B., \& Terquem, C. 2010, \mnras, 405, 573

\bibitem[Peale(1986)]{Peale1986}
Peale, S. J. 1986, in Satellites, ed. J. A. Burns \& M. S. Matthews
(Tucson, Univ. Arizona Press), 159

\bibitem[Peale(1999)]{Peale1999}
Peale, S. J. 1999, \araa, 37, 533

\bibitem[Petrovich et al.(2013)]{Petrovich2013}
Petrovich, C., Malhotra, R., \& Tremaine, S. 2013, \apj, 770, 24

\bibitem[Rivera \& Lissauer(2001)]{Rivera2001}
Rivera, E. J., \& Lissauer, J. J. 2001, \apj, 558, 392

\bibitem[Rein(2012)]{Rein2012}
Rein, H. 2012, \mnras, 427, L21

\bibitem[Shor(1975)]{Shor1975}
Shor, V. A. 1975, Celest. Mech., 12, 61

\bibitem[Sinclair(1978)]{Sinclair1978}
Sinclair, A. T. 1978, Vistas Astron., 22, 133

\bibitem[Tan et al.(2013)]{Tan2013}
Tan, X., Payne, M. J., Lee, M. H., Ford, E. B., Howard, A. W., Marcy,
G. W., Johnson, J. A., \& Wright J. T. 2013, \apj, submitted
(arXiv:1306.0687)

\bibitem[Tanaka et al.(2002)]{Tanaka2002}
Tanaka, H., Takeuchi, T., \& Ward, W. R. 2002, \apj, 565, 1257

\bibitem[Tittemore \& Wisdom(1990)]{Tittemore1990}
Tittemore, W. C., \& Wisdom, J. 1990, Icarus, 85, 394

\bibitem[Valencia et al.(2006)]{Valencia2006}
Valencia, D., O'Connell, R. J., \& Sasselov, D. 2006, Icarus, 181, 545

\bibitem[Ward(1997)]{Ward1997}
Ward, W. R. 1997, Icarus, 126, 261

\bibitem[Weiss et al.(2013)]{Weiss2013}
Weiss, L. M., et al. 2013, \apj, 768, 14

\bibitem[Wu \& Lithwick(2013)]{Wu2013}
Wu, Y., \& Lithwick, Y. 2013, \apj, in press (arXiv:1210.7810)

\bibitem[Xie(2012)]{Xie2012}
Xie, J. 2012, \apj, submitted (arXiv:1208.3312)

\bibitem[Yoder \& Peale(1981)]{Yoder1981}
Yoder, C. F., \& Peale, S. J. 1981, Icarus, 47, 1

\bibitem[Zhang \& Hamilton(2008)]{Zhang2008}
Zhang, K., \& Hamilton, D. P. 2008, Icarus, 193, 267

\end{thebibliography}
\end{document}